\documentclass[useAMS,usenatbib]{mn2e}
\usepackage{graphicx}

\title[Ruling out a massive-assymptoic giant-branch star as the progenitor of supernova 2005cs.]{Ruling out a massive-asymmptoic giant-branch star as the progenitor of supernova 2005cs.}
\author[J.J. Eldridge, S. Mattila  and S.J. Smartt]{J. J. Eldridge\thanks{E-mail: j.eldridge@qub.ac.uk}, S. Mattila  and S.J. Smartt.\\
Astronomy Research Centre, School of Maths \& Physics, Queen's University Belfast,Belfast, BT7 1NN, Northern Ireland, UK\\}

\begin{document}

\date{}

\pagerange{\pageref{firstpage}--\pageref{lastpage}} \pubyear{2002}

\maketitle

\label{firstpage}

\begin{abstract}
We calculate the predicted $UBVRIJHK$ absolute magnitudes for models of supernova progenitors and apply the result to the case of supernova 2005cs. We agree with previous results that the initial mass of the star was of low, around 6 to 8 M$_{\odot}$. However such stars are thought to go through second dredge-up to become AGB stars. We show that had this occurred to the progenitor of 2005cs it would have been observed in $JHK$ pre-explosion images. The progenitor was not detected in these bands and therefore we conclude that it was not an AGB star. Furthermore if some AGB stars do produce supernovae they will have a clear signature in pre-explosion near-infrared images. Electron-capture supernovae are thought to occur in AGB stars, hence the implication is that 2005cs was not an electron-capture supernova but was the collapse of an iron core.
\end{abstract}


\section{Introduction}

Core-collapse supernovae (SNe) are the spectacular events associated with the death of massive stars. They occur once the core is no longer supported by nuclear-fusion reactions or electron degeneracy-pressure. The core collapses to form a neutron star or, if the core is massive enough, a black hole. In the neutron star formation a large neutrino flux is produced that transfers a large fraction of energy to the stellar envelope that is ejected and produces the observed luminous display. 

There are two main evolutionary paths that lead to core-collapse. The first is the widely known iron core-collapse. This occurs in stars initially more massive than about 10 M$_{\odot}$\footnote{The value varies with the details of convection in stellar models. Here we quote masses using convective overshooting.} where nuclear burning progresses all the way to production of an iron core. Iron being the most stable element does not provide energy by further fusion reactions so the core collapses. The second core-collapse path is restricted to stars below about 10M$_{\odot}$. In these stars nuclear burning progresses no further than carbon burning. After carbon burning neutrino cooling reduces the temperature of the core preventing further nuclear reactions. Core collapse occurs when the oxygen-neon (ONe) core reaches the Chandrasekhar mass (M$_{\rm Ch} \approx 1.4M_{\odot}$) and electron degeneracy-pressure can no longer support the core. The central density increases until a density of around 10$^{9.6}$ g/cm$^{3}$ is reached. Then electron-capture by magnesium-24 ($^{24}$Mg) and/or neon-20 ($^{20}$Ne) reduces the electron degeneracy-pressure further and the collapse accelerates \citep{ecapture1,emore2}.

Electron-capture SNe are thought to occur predominately in massive Asymptotic Giant-Branch (AGB) stars \citep{WHW02}. AGB evolution occurs after helium burning in stars from 0.8 to about 8 M$_{\odot}$. After the main sequence and formation of a helium core a star will ascend the red giant branch. When this occurs the convective envelope penetrates into the helium core mixing products from hydrogen burning to the surface, this process is known as dredge-up. It decreases the hydrogen, carbon and oxygen abundance while increasing helium and nitrogen abundance. After some time helium will ignite in the core and the star will move back to the blue. When helium burning ends and a carbon-oxygen (CO) core is formed the star moves back to the giant branch and dredge-up occurs for the second time. This time dredge-up penetrates deeper and mixes up nearly all the helium leaving just a thin layer covering the CO core and the hydrogen and helium burning shells in close proximity. This arrangement is unstable and nuclear burning progresses as a series of pulses. The hydrogen burning shell burns until the layer of helium is thick enough to ignite. This helium rapidly burns until it is exhausted and the cycle restarts. AGB stars are more luminous than red giants of the same mass because the hydrogen shell is at a much higher temperature in close proximity to the helium burning shell. In the most massive AGB stars core carbon-burning occurs around the time of 2nd dredge-up and so there is an ONe core (carbon burning extinguishes after the CO core is completely converted to an ONe core and never reignites). This means that if the core grows via thermal pulses it may reach the M$_{\rm Ch}$ and produce an electron-capture SN. It is a race between mass-loss and core growth as to whether a SN will occur. The evolution of these stars has been well studied, e.g. \citet{ecapture2,sagb3} and \citet{sagb4}. While \citet{ETsne} and \citet{tagb} have concentrated on the mass-ranges over which these events occur and when these SNe might occur.

It is surprising that two quite different mechanisms for producing a neutron star could lead to similar type IIP SN events. Type IIP SNe are classified by having hydrogen lines in their spectra and a long plateau phase in their lightcurves during which the luminosity is roughly constant. To produce this type of SN a massive and extended hydrogen envelope is required. This envelope which produces the SN display does not differ substantially between the two paths of core-collapse. Discriminating between these two SN types can only be achieved by either observing differences in the nucleosynthesis products from the different collapse mechanisms, or by observing the progenitor star. In this letter we demonstrate the latter method for SN 2005cs. 

During the SN explosion nucleosynthesis occurs and one product is nickel-56 ($^{56}$Ni) from explosive burning of oxygen and silicon \citep{WHW02}. $^{56}$Ni provides the late time SN luminosity as it decays to cobalt-56 ($^{56}$Co) and then iron-56 ($^{56}$Fe). SN 2005cs produced a lower than average amount of nickel in the explosion, of the order of 0.01 M$_{\odot}$ \citep{andrea,ni56t}. Pastorello et al. (in prep) who use image subtraction to remove the background to accurately estimate the sources late-time SN luminosity. They suggest it could be as low as 0.004 M$_{\odot}$. Typical nickel masses are for example 0.016 M$_{\odot}$ for SN 2003gd \citep{03gd} or 0.075 M$_{\odot}$ for SN 1987A. To produce nickel, oxygen and silicon are required. Low nickel mass therefore indicates that there was very little of these two elements around the collapsing core and such a structure occurs within massive-AGB stars. Thus the progenitor of 2005cs is a good candidate to be such a star. Furthermore simulations of collapsing ONe white-dwarfs indicate that they produce less than 0.001 $M_{\odot}$ of $^{56}$Ni \citep{aic}.

In this letter we first discuss our synthetic photometry method to calculate $UBVRIJHK$ magnitudes for our stellar models. We then compare model predictions to the observational progenitor limits for SN 2005cs. It was detected in a pre-SN $I$ band image but not in deep $JHK$ observations. We show that the latter can provide a strong limit on whether 2nd dredge-up had occurred

\section{Synthetic $UBVRI$ \& $JHK$ magnitudes.}

Stellar evolution codes only produce a few details of the observable characteristics such as the luminosity, radius, surface temperature and composition where the surface is defined to be where the average optical depth reaches two thirds. Comparing models to observed stars can only be done accurately from spectroscopic observations of the stars. However spectroscopic data are generally not available for SN progenitors, apart from the rare case of SN 1987A. Recent work in searching for information on the progenitors of supernovae has focused, successfully, on finding observations through numerous broad-band filters (combinations of $UBVRIJHK$).  Calculating stellar parameters from broad-band photometry obviously provides less accurate results than from spectroscopic analyses, but the method is the only viable one open to us unless a Milky Way, or possibly Local Group SN occurs. Previous attempts to determine stellar parameters of SN progenitors assigned bolometric corrections based on stellar type (effective surface temperature) for the progenitors estimated from broad-band colours e.g. \citet{SMARTT}. However an alternative method is to predict the photometric magnitudes of stars along theoretical stellar evolution tracks using atmosphere models. 

\begin{figure}
\includegraphics[angle=0, width=84mm]{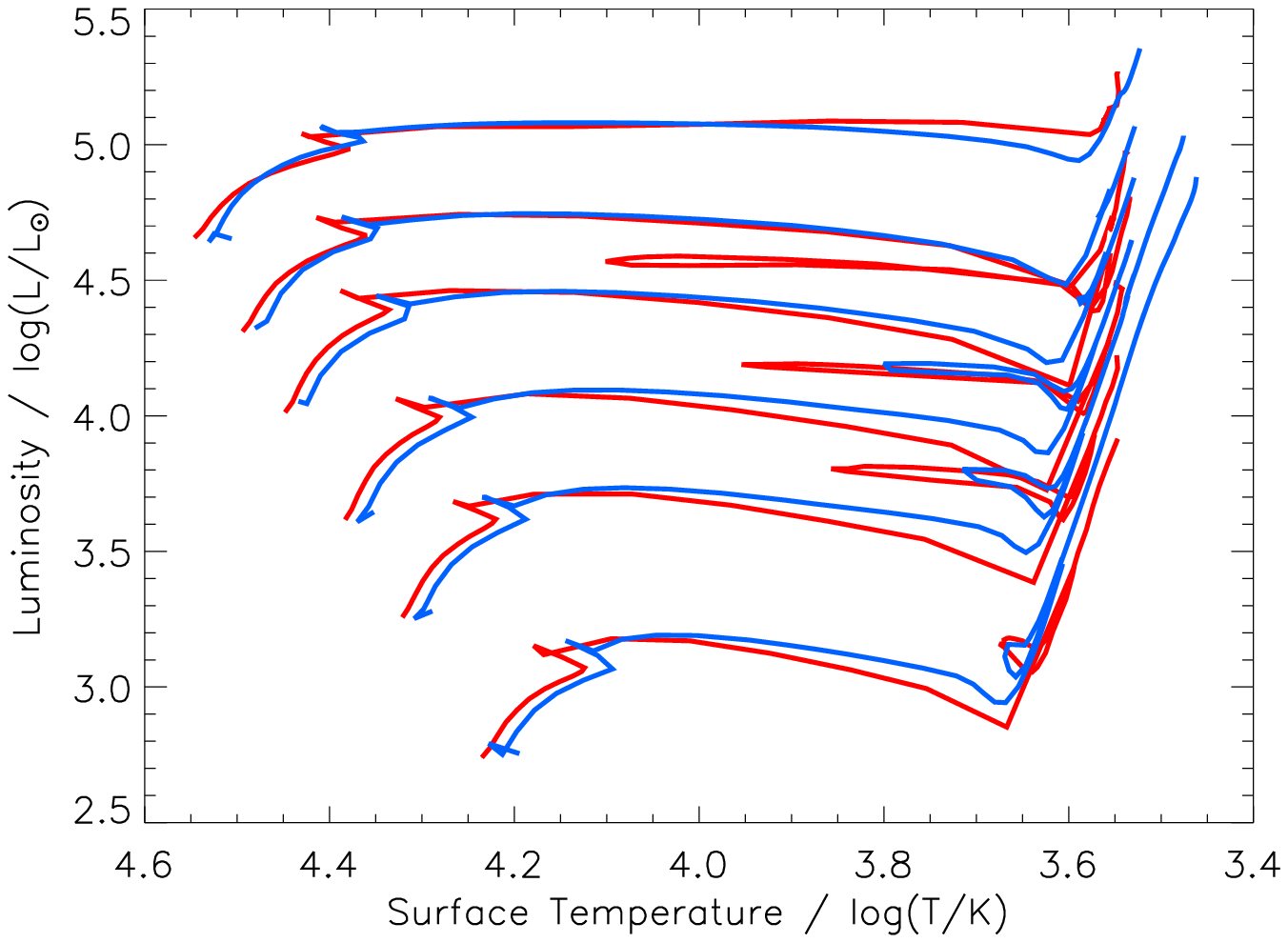}
\includegraphics[angle=0, width=84mm]{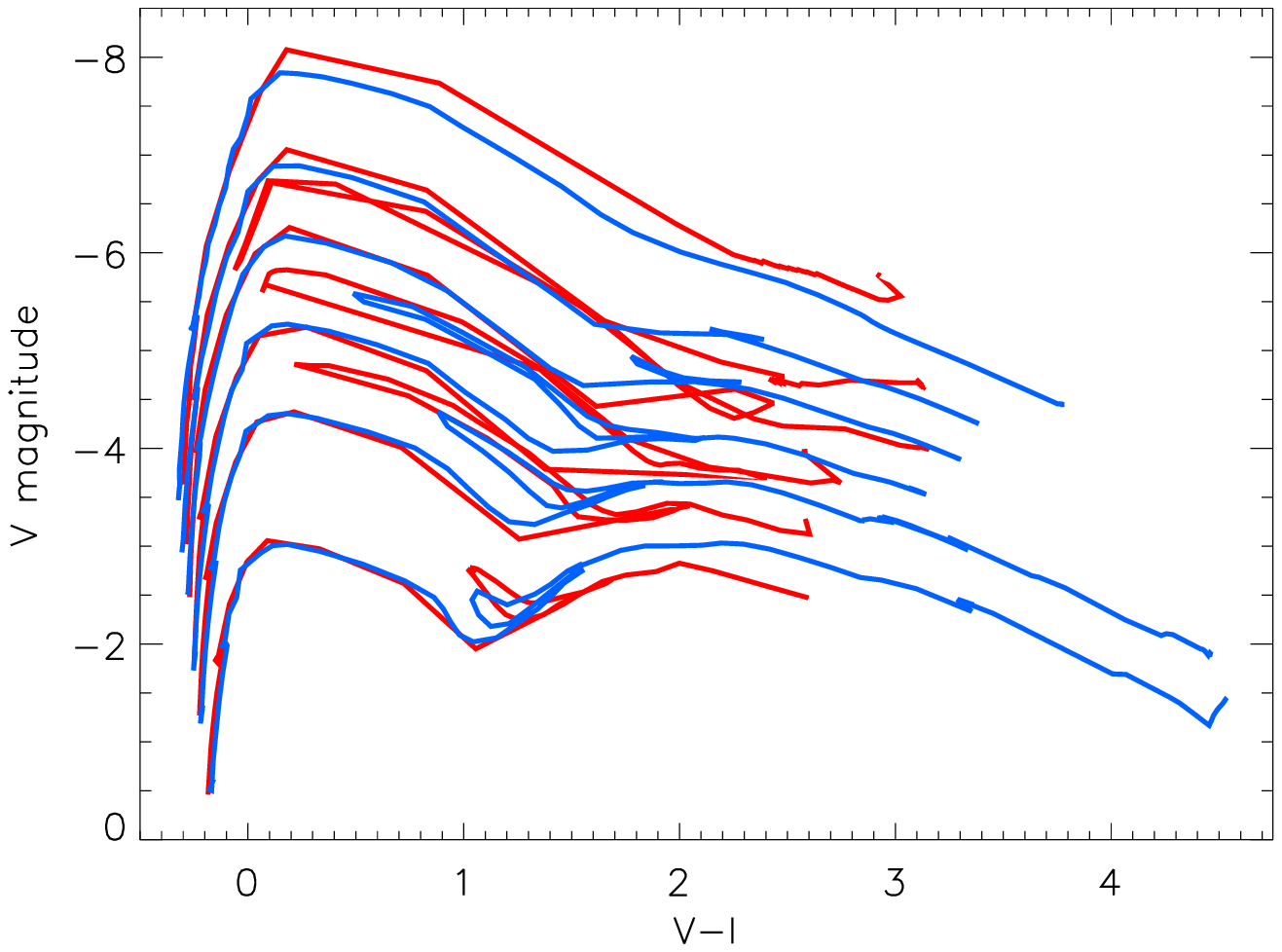}
\caption{Hertzsprung-Russel (HR) diagrams comparing the STARS models (red lines) with the Geneva models (blue lines). From bottom to top the initial masses are 5, 7, 9, 12, 15 and 20M$_{\odot}$. Upper panel is plotted with $\log(L/L_{\odot})$ versus $\log(T_{eff}/K)$ and the lower panel is plotted by the absolute V magnitude and the V-I colour.}
\label{hrO}
\end{figure}

We used the method of \citet{LS01} to perform synthetic photometry on the Cambridge STARS evolution models \citep{E1} (see \texttt{http://www.ast.cam.ac.uk/$\sim$stars} for more details). This involves using the BaSeL model atmosphere library \citep{basel3} to work out the flux in different broad-band filters \citep{ubvribands,jhkbands} for different surface temperatures, gravities and metallicities. Then interpolating in this grid to obtain colours for each model. To calculate the absolute magnitudes we used a theoretical spectrum of Vega from the same atmosphere library, setting all colours to zero. In this we assumed a radius of 3.1 R$_{\odot}$ for Vega which is larger than its known radius of $2.7 \, R_{\odot}$ \citep{vega}. However using the theoretical Vega spectrum it provides the correct flux for Vega in the $V$ band. The adopted radius is the primary source of error in the absolute magnitudes but does not affect the colours.

The resultant colours agree with the colours from the commonly used Geneva models \citep{LS01}. Both sets of models are plotted in theoretical Hertzsprung-Russel diagrams in Figure \ref{hrO}. The slight disagreement is due to differences in the two evolutionary codes, such as updated opacity tables, different initial abundances and different mass-loss rates. Also our models extend further in time as they are calculated to the beginning of neon burning while the Geneva models end after core carbon-burning and do not have 2nd dredge-up. This means the track endpoints of our models are more luminous and slightly cooler. Our models for 8 M$_{\odot}$ and below experience 2nd dredge-up.

It is difficult to estimate the uncertainty in our colours. The largest errors are due to the assumption of the radius for Vega and also assuming that all filters have an absolute magnitude of zero for Vega. We estimate the error in the absolute magnitudes due to this are around $\pm$0.3 mags. This was estimated by using different methods to calibrate the zero point of our system such using the colours of Sirius and the Sun, and varying the radius of Vega.

\begin{figure}
\includegraphics[angle=0, width=84mm]{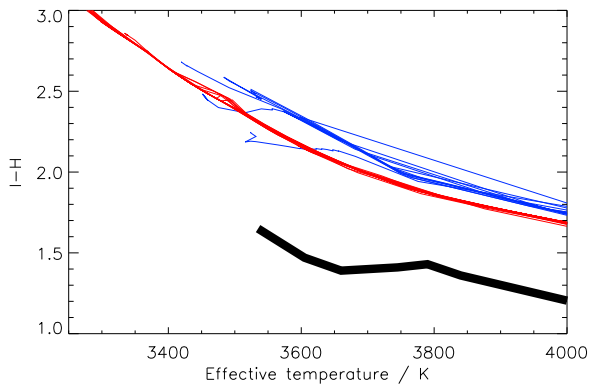}
\caption{The comparison of observed and predicted $I-H$ colours for RSGs. The thick black lines are the observed colours from \citet{obsrsgs}. The red lines are the STARS models, while the blue lines are from the Geneva models.}
\label{IH}
\end{figure}

We have checked the predicted colours against those of observed RSGs by \citet{obsrsgs}. They supply the average colours expected for different spectral types of RSGs. However recently there has been a reassessment of the effective temperatures of RSGs by \citet{cool} which we take into account in our analysis. We show examples of the comparisons in Figure \ref{IH}. The agreement of the models with the data of \citet{obsrsgs} is poor with the average discrepancy of around 0.5 to 0.7 mags for all the colours. We have also calculated the expected colours for RSGs from the MARCS atmosphere models which were used by \citet{cool} to derive their temperature calibration. We find that the colours are close to those from the atmosphere models we use with a discrepancy of less than 0.1 mags.

It is more difficult to compare the predicted colours of the AGB stars with observations. While lists of AGB colours exist \citep{agbcol}. AGB stars come from a wide range of initial masses ($0.8 \la M \la 7 M_{\odot}$) and therefore have a wide range of core masses and luminosities. However the AGB stars with larger cores, those closest to the point of core-collapse, tend to be the most luminous one. The predicted colours of our AGB stars broadly agree with the observed AGB colours of \citet{agbcol} and \citet{agbfluks}. Also we have compared the colours to the more detailed AGB atmosphere models of \citet{agbgro} and again there is a broad agreement but the range of possible colours is much greater in these samples. We will use these AGB colours as well as those from our models to determine whether the progenitor of SN 2005cs was an AGB star.

The disagreement between the observed and theoretical colours is in general greatest at the end points of the stellar models where the RSGs are most luminous and extended. This may be due to the way the stellar radius is calculated in theoretical stellar models. Stellar evolution codes use wavelength averaged opacity and the photosphere is defined to be where the optical depth becomes two thirds. However in RSG atmospheres opacity is wavelength dependent which results in longer wavelength (infra-red) emission coming from a smaller radius than the shorter wavelength (visible) emission \citep{bj}. If this radius is not comparable to RSG radii then a systematic error is introduced to the predicted colours. However there is little we can do to correct the models but we can calibrate our predicted colours by introducing an empirical correction. Therefore we produced a second set of tracks where for RSGs we correct the $JHK$ magnitudes to agree with \citet{obsrsgs}. The corrections we adopt are: $J_{\rm c}=J+0.5$, $H_{\rm c}=H+0.5$ and $K_{\rm c}=K+0.7$.

\section{Implications for the progenitor of SN 2005cs.}

\begin{figure}
\includegraphics[angle=0, width=84mm]{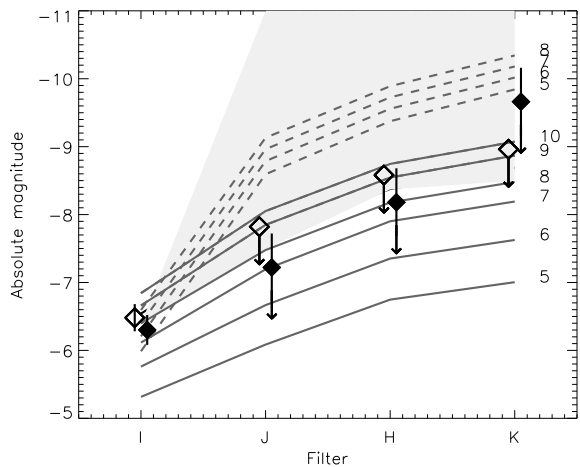}
\caption{The predicted progenitor magnitudes from the STARS code, with the correction applied, compared to the observed $I$ band detection and $JHK$ limits. The empty diamonds are from \citet{Maund} and the filled diamonds from \citet{Li}. The shaded region is defined by the maximum and minimum colours for AGB stars from \citet{agbgro}. The solid lines are for RSG models while the dashed lines are for AGB models.}
\label{starstwo}
\end{figure}

The progenitor of SN2005cs was detected in deep Hubble Space Telescope (HST) $I$ band images of M51 taken in January 2005, a few months before the SN occurred in June 2005. In $BVR$ and $JHK$ filters the progenitor was not detected. Two studies exist that use the same HST image to calculate $I$ band magnitudes for the progenitor but use different near infra-red data. The $BVR$ upper limits indicate that the progenitor could not have been a blue supergiant and was a red supergiant no hotter than a K5Ia type \citep{Li,Maund}. However the  $JHK$ limits also constrain how cool the progenitor was. \citet{Maund} made use of ground-based Gemini observations taken in April 2005 to produce limits on the $JHK$ magnitudes while \citet{Li} used HST NICMOS observations taken in June 1998 to produce deeper $JH$ but shallower $K$-band limits on the progenitor. Both studies used the observations of \citet{obsrsgs} to limit the spectral type of the progenitor star.

The non-detections in the $JHK$ bands can be used to place very sensitive limits on the luminosity and temperature of the progenitor. If we decrease the surface temperature of the progenitor while maintaining the I band magnitude we must increase the progenitor's luminosity and the synthetic $JHK$ magnitudes become greater than the upper limits. This restricts the type of progenitor severely.

In Figure \ref{starstwo} we plot the $IJHK$ magnitudes of the STARS models with the correction. In calculating the magnitudes we have used a distance modulus of 29.62 \citep{m51dis}. Over these lines we then add the $I$ band detection and the $JHK$ limits from both \citet{Maund} and \citet{Li}, corrected for dust extinction using an $A_{\rm V}$ of 0.34 \citep{andrea} and the extinction law of \citet{dust}. Without the correction it is not possible to fit the $I$ detection and still stay below the $JHK$ limits unless we adopt the very upper ends of the error bars of the \citet{Li} as their limits. With the correction to the $JHK$ colours applied in Figure \ref{starstwo} it is easier to fit the $I$ band detection and the near-infrared upper limits. This supports our method of correcting our model synthetic photometry to match the observed JHK colours of \citet{obsrsgs}.

The most important feature to notice is that if the progenitor had gone through second dredge-up to become an AGB star, its colours would be as shown by the dashed lines and it would have been clearly detected in the near-infrared bands. This is made more certain by the shaded region showing the range of AGB colours from \citet{agbgro}. Most of the AGB colours are far above the $JHK$ limits but those with the lowest values for the $JHK$ magnitudes have very low mass-loss rates and there are many more models with higher values. Therefore the conclusion that the progenitor was not an AGB star is firm.

For the STARS and Geneva models the progenitor would be around 6 to 8 M$_{\odot}$. This of course agrees with the conclusions of \citet{Maund} and \citet{Li}. But now we can limit the amount of 2nd dredge-up that occurred in the progenitor. We can conclude that the progenitor did not experience the extreme dredge-up that produces an AGB star. However it is possible that the progenitor did experience a small amount of dredge-up decreasing the helium core mass slightly, by a few tenths of a solar mass. This is predicted in some of the models shown by \citet{tagb} and this would slightly decrease the luminosity of the progenitor meaning we would underestimate its initial mass. Therefore a more robust estimate of the final helium core mass for the progenitor is between 1.7 to 2.2 M$_{\odot}$.

The important result is that the 2nd dredge-up did not lead to an AGB star. If it did it would be clearly detected in every $JHK$ image available while agreeing with the I band detection. Therefore this sets an upper limit for the masses of AGB stars and also AGB stars as SN progenitors of 6 to 8 M$_{\odot}$. This overlaps with the upper limit for AGB stars to produce a white dwarf of around 6.8 to 8.6 M$_{\odot}$ from \citet{wd}.

\section{Discussion of uncertainties.}

The uncertainties in our estimation of the progenitor mass come from three sources. Firstly errors in the observational photometry, secondly errors in the synthetic photometry and third errors due to the lack of understanding of RSGs.

The $I$-band photometric-measurements of \citet{Maund} and \citet{Li} differ by 0.3 mags. This is mainly due to the different methods employed to remove the flux from the nearby bright objects. However we can see in Figure \ref{starstwo} a difference of 0.3 mags in the $I$ band actually makes little difference to the conclusions we draw. As discussed above, it is the $JHK$ limits that constrains the progenitor more severely, particularly cool AGB stars.

There are four uncertainties in the synthetic photometry. The first two are the assumptions of the Vega zero-point for all colours and our assumed Vega radius as mentioned above that introduces an error of around 0.3 dex. Third, the use of the filter functions for the $UBVRIJHK$ bands; the observed $I$ band magnitude was in fact estimated from a HST filter which is quite different from the Johnson filters. However we have compared the colours predicted by the different $JHK$ filter functions and find that the difference is typically less than 0.05 dex. We also assumed that the synthetic spectra are accurate, however any error in their calculation will result in incorrect magnitudes. For example the models do not include mass-loss. Another uncertainty arises from the use of a wavelength averaged optical depth to calculate the stellar radii as discussed above. Other uncertainties in the stellar evolution code, such as use of mixing length theory to describe convection could affect our model parameters. However these are accounted for in our correction factors.

Uncertainty is also present due to a number of AGB star and RSG enigmas; dust, oscillations and asymmetry. In this case dust is not an important uncertainty. Any dust extinction in the $IJHK$ bands is much less than the $V$-band \citep{dust}. Because our conclusions only depend on these bands they are not affected by dust extinction unless a very large reddening (of order $A_{v} \sim 10$) is invoked. But this is inconsistent with all estimated of reddening towards this SN \citep{Maund,Li,andrea,baronie}.

A larger problem is whether the progenitor was oscillating or pulsating before the explosion. \citet{pulse} performed calculations to follow the oscillations a red supergiant may undergo before explosion. They show large changes in the surface details of the star. However there is no observational evidence how a star might oscillate before a SN. We have included by hand oscillations in our stellar models of a similar magnitude as those of \citet{pulse}. We find that while the $I$ band magnitude can vary a great deal the $JHK$ magnitudes vary only slightly.

The last problem of asymmetry is best shown by observations of Betelgeuse \citep{bj}. Not only does this work show that the radius of a RSG is wavelength dependent but we also see that the near infra-red light comes from deeper in the star than the visible light. This visible light, originating from nearer the surface may be more affected by star spots and this would make the star look more luminous in the $I$ band than the $JHK$ bands.

These many different uncertainties do contribute to the error on the estimation of the initial mass but do not affect our conclusion that the progenitor did not experience 2nd dredge-up and was not an AGB star, arguing against the core-collapse mechanism being electron capture for SN 2005cs. It is worthwhile to compare the progenitor mass for this SN to the upper mass limit for an AGB star from \citet{wd}. It is striking that the masses match at around 6 M$_{\odot}$. Therefore studies such as \citet{tagb} are required to more deeply understand this important change over in behaviour from an AGB star to a SN progenitor.

In conclusion the progenitor of SN 2005cs was a low mass star of between 6 and 8 M$_{\odot}$ and may have experienced slight 2nd dredge-up but not a large amount of dredge-up that would have produced an AGB star. Furthermore if AGB stars are the progenitors of some SNe they will leave a clear signature in near-infrared pre-explosion images. An attempt to put limits on SN progenitors in the near infra-red in a systematic way is already underway by our group with deep imaging of about 50 galaxies in $JHK$ with VLT, ISAAC, Gemini and UKIRT \citep{Maund}.

\section*{Acknowledgments}
This work, conducted as part of the award ``Understanding the lives of massive stars from birth to supernovae'' made under the European Heads of Research Councils and European Science Foundation EURYI Awards scheme, was supported by funds from the Participating Organisations of EURYI and the EC Sixth Framework Programme. Also JJE would like to thank Norbert Langer  and Andrea Pastorello for useful discussion.

\bsp

\label{lastpage}

\end{document}